# Printed Circuit Board Metal Powder Filters for Low Electron Temperatures

Filipp Mueller[1,a], Raymond N. Schouten[2], Matthias Brauns[1], Tian Gang[1], Wee Han Lim[3], Nai Shyan Lai[3], Andrew S. Dzurak[3], Wilfred G. van der Wiel[1], Floris A. Zwanenburg[1]

[1]*NanoElectronics Group, MESA+ Institute for Nanotechnology, University of Twente, P.O. Box 217, 7500 AE Enschede, The Netherlands*

[2]*Kavli Institute of Nanoscience, Delft University of Technology, 2600 GA Delft, The Netherlands*

[3]*ARC Centre of Excellence for Quantum Computation and Communication Technology, The University of New South Wales, Sydney 2052, Australia*

We report the characterisation of printed circuit boards (PCB) metal powder filters and their influence on the effective electron temperature which is as low as 22 mK for a quantum dot in a silicon MOSFET structure in a dilution refrigerator. We investigate the attenuation behaviour (10 MHz- 20 GHz) of filter made of four metal powders with a grain size below 50 μm. The room-temperature attenuation of a stainless steel powder filter is more than 80 dB at frequencies above 1.5 GHz. In all metal powder filters the attenuation increases with temperature. Compared to classical powder filters, the design presented here is much less laborious to fabricate and specifically the copper powder PCB-filters deliver an equal or even better performance than their classical counterparts.

## I. INTRODUCTION

In analogy to natural atoms where optical methods are used to probe the discrete energy spectrum, electron transport spectroscopy has proven to be a very powerful method for mapping out the corresponding levels of quantum dots, also known as "artificial atoms". As is generally true for spectroscopic analyses, it is desired to minimize extrinsic sources of broadening of resonances for accurate electron spectroscopy. The FWHM of a Coulomb resonance in a quantum dot is set by the quantum mechanical level broadening $\hbar\Gamma$ (determined by the tunnel coupling between quantum dot and electron reservoirs) and the effective electron temperature $T_e$ (thermal broadening of Fermi-Dirac distribution) [1]. This thermal broadening has contributions from the bath temperature $T_{bath}$ of the cryogenic setup and additional inelastic processes originating from noise and interference of the measurement environment.

In electronic transport experiments $T_e$ can easily be an order of magnitude higher than the cryostat's base temperature $T_{bath}$. The latter can be as low as ~10 mK in dilution refrigerators [2]. At those temperatures the elevation of $T_e$ is mainly caused by electrical noise and the pick-up of interference, which exists in the entire frequency spectrum. Effective filtering is thus crucial for the effective electron temperature to approximate the base temperature as close as possible. Different types of filters are used for different frequency ranges. RC-filter cover the low-frequency range [10 Hz – 10 MHz], Pi-filters intermediate frequencies [10 MHz – 1 GHz] and since their invention in 1987 by Martinis *et al.* [3], metal powder filters are commonly used as microwave absorbers [100 MHz – 100 GHz].

---

[a] Author to whom correspondence should be addressed. Electronic mail: f.muller@utwente.nl.

## II. METAL POWDER FILTERS

Metal powder filters are fabricated as follows: a metallic wire of ~1-2 m length is wound around a metal rod which is mounted in a tube and finally the tube is filled with metal powder. According to Martinis *et al.* the attenuation in those filters is caused by skin-effect damping [3,4]. Alternating electric currents lead to eddy currents in a conductor due to corresponding alternating magnetic field. Thus, the current mainly flows near the surface of the conductor, within the so-called skin depth, and increase the effective resistance. Because metal powder has an extremely large surface compared to bulk, one expects an enormous dissipation due to the skin effect. Usually, a mixture of epoxy and metal powder is used to remove the dissipated heat from the grains to the environment. The skin depth depends on the AC frequency $f$, the metals resistivity $\rho$ and magnetic permeability $\mu$ according to $\delta = \sqrt{\rho / \pi \cdot f \cdot \mu}$. The effective resistance and thus the dissipation increase with frequency, making powders very efficient absorbers of high frequency signals. This has been shown in earlier experiments. Lukashenko *et al.* [5] showed that the attenuation of their copper powder filter reached an attenuation level of –90 dB at 6.2 GHz whereas the stainless steel powder filter did so already at 1.0 GHz. Besides the higher resistivity of stainless steel at 4 K the higher magnetic permeability of stainless steel will make δ even smaller, thus increasing the dissipation. Fukushima *et al.* [6] showed that both stainless steel and copper powder attenuate more at room-temperature than at 4.2 K. The change in attenuation after cooling down was smaller for stainless steel powder than for copper, likely due to the difference in residual resistance ratios of the two metals. This has motivated us to investigate different metal powders, e.g. manganin, a metal compound commonly used as low-temperature wiring because of its high resistivity and near-zero temperature coefficient of resistance.

In this work we present a novel type of metal powder filter using printed circuit boards as building block. Figure 1(a) shows one side of the used printed circuit board (PCB). The substrate material is 1.55 mm thick standard FR4, low-cost and lossy at microwave frequencies. PCB tracks are 100 μm wide and have a thickness of 35 μm copper covered with 100 nm gold. The PCB has a length of 128 mm and contains 12 printed wires, six on each side. Each wire has a length of 120 cm where two neighbouring wires are meandered in a double S shape. The PCB thickness and the 12 solder pads on top and bottom are chosen to solder 0.2 inch pin headers that makes it an easy plug-and-play system in measurement setups. Here, SMA connectors are used to test the filter characteristic at high frequencies (Fig. 1 (c)). The solder joints are covered with a layer of epoxy (6305950 -Bison Kombi Snel®) for a better impedance matching. Beside the bare PCB different metal powders were investigated to compare the influence of the metal powder on the attenuation behaviour:

copper, brass (Cu 68.5-71.5%, Pb 0.07% max, Fe 0.05% max, Zn remainder), stainless steel (alloy 302/304) and manganin (Cu 86%, Mn 12%, Ni 2%) with a grain size below 50 μm. Finally, a 3 mm thick layer of metal powder/epoxy mixture (50:50 in volume) is glued on each side of the PCB. In addition, the small distance between the paired wires (100 μm) reduces the electromagnetic pick-up and thus pursues a twisted pair like wiring (signal and ground) which is commonly used in low-noise measurements setups.

In conclusion, this metal powder filter's design is more flexible and reproducible. It hosts many parameters that can be changed and investigated to maybe increase and optimize the filter's attenuation and impedance characteristic, e.g. width and thickness of the lines, distance between the lines, length of the PCB etc.. In addition, the fabrication of the filters is much less time-consuming in mass production.

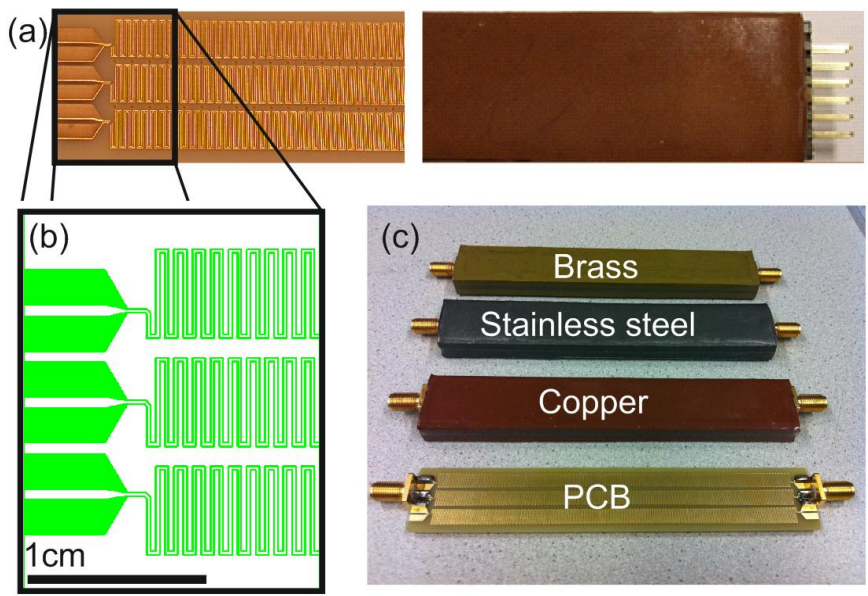

FIG. 1. Metal powder filter layout  (a): left: one side of the printed circuit board (PCB) with three paired lines; right: PCB with soldered pin-headers and copper powder/epoxy mixture. Length of each line is 1.2 m. (b): in-zoom to one of the ends. Six lines end up in a solder pad for 0.2 inch pin-headers. (c): powder filters with different metal powders with a grain size below 50 μm; a 3 mm thick layer of epoxy/powder mixture (50:50 in volume) on each side of the PCB. Here, SMA connectors were used to perform high-frequency measurements.

### III. ATTENUATION CHARACTERISTICS
#### A. Room temperature

The filter attenuations were measured using a vector network analyser (VNA) Rohde&Schwarz ZVB20 (10 MHz-20 GHz) with a resolution bandwidth set at 10 Hz. Standard calibration procedure is used to extract the attenuation of the high frequency lines from the filter characteristic. Figure 2(a) shows the transmission characteristic of a copper powder filter for different configurations at room temperature. Plotted is the attenuation of the transmitted signal in decibel (dB) versus frequency. At ambient conditions, the copper powder filter reaches an attenuation level of -80 dB at 2 GHz. Above 2 GHz the transmission exhibits an upturn to -60 dB which is suppressed when surrounding the copper powder filter with a layer of Eccosorb® MCS [7], a high loss microwave absorber. The transmission is suppressed even to the noise floor above 4 GHz when inserting the copper powder filter into a closed copper tube where the residual space is filled up with Eccosorb (see inset Fig. 2(a)). Measurements of the last configuration for the bare PCB, brass- and stainless steel powder filter are shown in Fig. 2(b). All curves display typical attenuation behaviour. Note that already the bare PCB exhibits an attenuation of -80 dB at 7 GHz. Resonances appear in the transition region between pass band and stop band. While stainless steel reaches -80 dB at 1.5 GHz, copper and brass reach this attenuation at a slightly higher frequency of 2 GHz.

The increasing signal at higher frequencies of the copper powder filter under ambient conditions can be explained by pick-up of external electro-magnetic radiation. Using a closed copper tube which serves both as housing for the powder filters and as Faraday-cage highly attenuates the signal by about 30 dB. The origin of the resonances in the transition region is not clear. Since powder filters are low pass filters in the microwave range ($f_{cut\text{-}off}$ ≈ 100 MHz- 1 GHz), good filters attenuate high frequency noise by more than 80 dB. The lower the cut-off frequency and the steeper the fall-off in the transition range the better the filter. While all filters exhibit almost the same slope stainless steel powder shows the best performance. The performance of the bare PCB is explained by the fact that the paired layout (signal+ ground) forms a lossy transmission line. This improves the filter action because high frequency interference is not reflected at the input of the filter but absorbed inside. This impedance control is an improvement compared to other multi-wire solutions.

We conclude that stainless steel powder is favourable at room temperature. However, for quantum low-noise measurements we search for the best filter at low temperature. We tested all filters in a cryogenic setup (Oxford Triton 200) both at room temperature and 10 mK, for comparison.

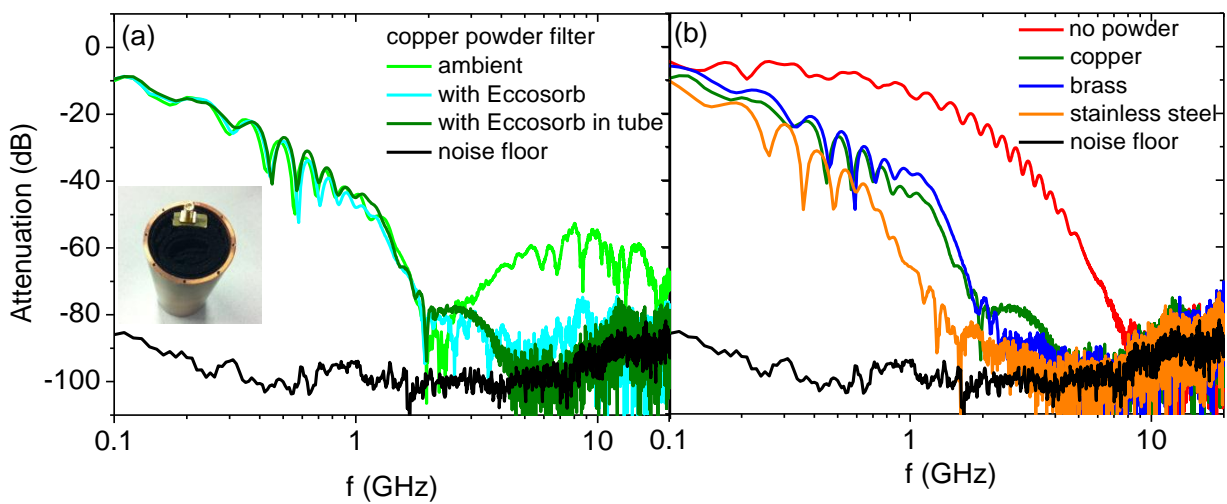

FIG. 2. Frequency response of different powder filters at room temperature. The black curve represents the noise floor of the vector network analyser. (a): Copper powder filter in different configurations. Inset: powder filter in a copper test tube filled with Eccosorb. (b): Frequency response of different metal powders filters and the bare PCB.

**B. Temperature dependence**

The setup for the two-port measurements at low temperature is shown in Fig. 3(a). Port1 and Port2 at the VNA are connected to the high frequency lines in the Triton via 1.5 m radio frequency cables (GORE® PHASEFLEX® Microwave/RF Test Assemblies) which are specified up to 26.5 GHz. To avoid direct thermal connection between room temperature and milliKelvin temperature via the high frequency lines, three times 3 dB attenuators (Aeroflex Weinschel, fixed coaxial attenuator, model: 4M-3, DC-18 GHz) in each high frequency line ensure thermal anchoring to plates with different temperatures. The copper can, hosting the filters, is mounted at the mixing chamber plate reaching a base temperature of 10 mK. Residual space in the can is filled with Eccosorb. Flexible radio frequency cables (LEAD, RG316 SMA M/M, 0.5m, DC-3 GHz) connect the filters to the high frequency lines.

The transmission characteristic of different filters in this setup at room temperature and cryogenic temperature are plotted in Fig. 3(a) and (b), respectively. Moreover the throughput of the high frequency lines and the noise floor after calibration (HP calibration kit 85052B) are shown. At frequencies higher than 6 GHz the throughput signal becomes slightly noisy and the noise floor increases because of signal losses. We reproduced the attenuation curves at room temperature from Fig. 2(b) in our cryogenic setup, but with higher noise level due to the attenuators. Beside different metal powder filters the measurement data in Fig. 3(a) also includes the bare PCB where the metal powder part is substituted by vacuum space and a layer of Eccosorb, respectively. As one see the Eccosorb layer mainly suppresses the resonances in the transition regime. Figure 3(b) shows measurements at cryogenic temperature. Compared to room temperature all filters show a transition range shifted to higher frequencies. We observe that stainless steel and manganin begin to filter out at slightly higher frequencies than copper and brass but show steeper fall-off.

We attribute the increasing noise floor and noisy through-put at frequencies above 6 GHz to the loss in the used flexible RF cables which are specified up to 3 GHz. Housing the bare PCB in a Faraday cage, high frequency cavity modes occur. Eccosorb directly attached to the PCB clearly damps these modes. A significant shift of the transition range to lower frequencies indicates the effect of the metal powder on the attenuation.

Literature suggests that the filter performance depends on the electrical resistance (energy los via skin effect at surface of particles) of the metal [4,5,8]. While the resistivity of all metals at room temperature are quite comparable (in the order of $10^{-8}$ Ωm) the magnetic property could explain the better filter working of stainless steel powder at room temperature. Considering that the skin effect would be the reason for the signal absorption, we expect different transmission characteristics of the filters at cryogenic temperatures according to their resistivity at low temperatures. Although the difference in resistivity e.g. between copper and stainless steel is about 3 orders of magnitude, we surprisingly found that all metal powder filters show comparable filter characteristic at 10 mK (Fig. 3(b)), in contrast to our expectations.

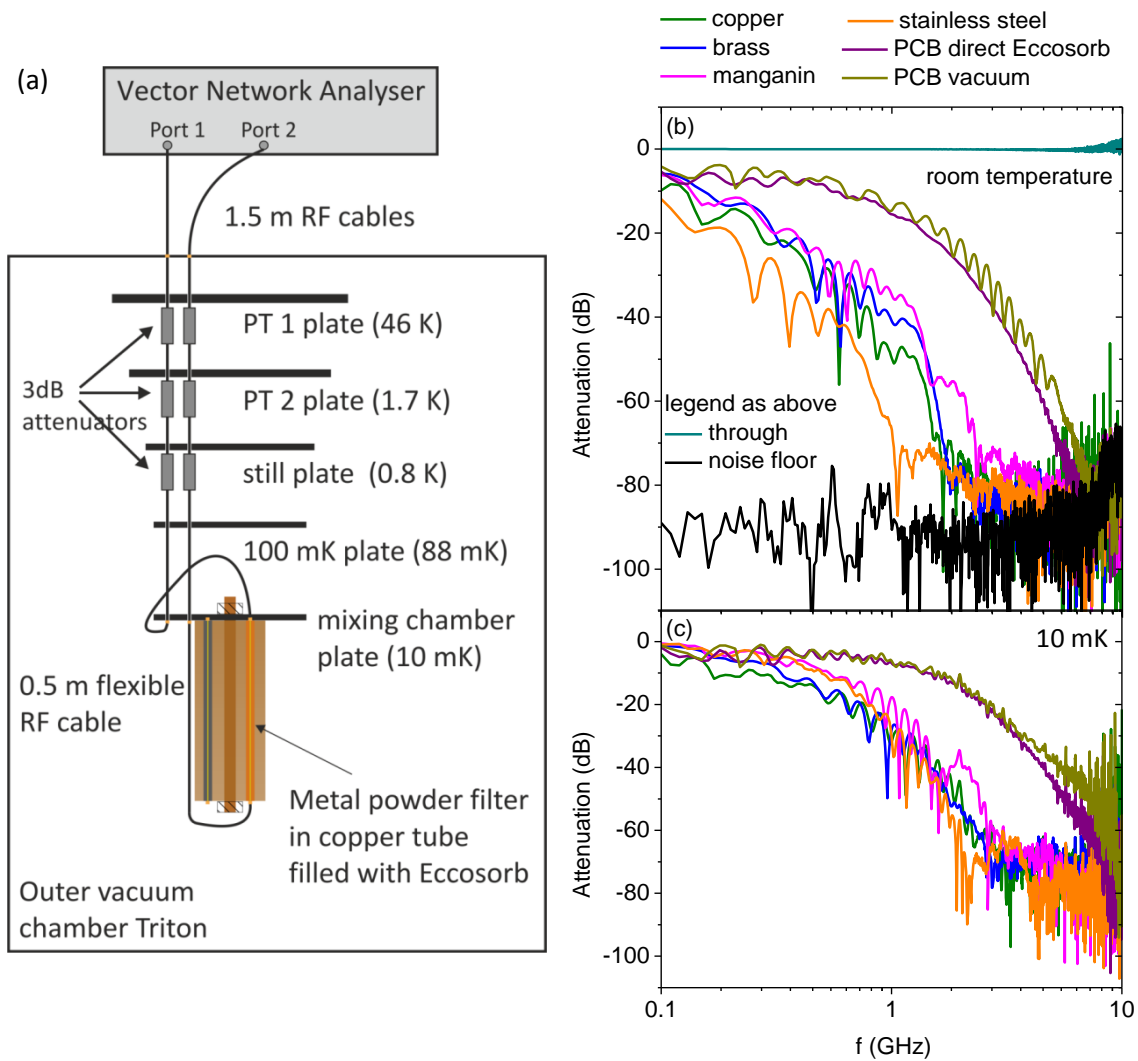

FIG. 3. Measurement setup and transmission characteristics of different metal powder filters both at room- and cryogenic temperature (a): Cryogenic measurement setup. Three 3dB attenuators ensure thermal anchoring of the inner conductor of the high frequency lines to different temperature stages. The copper can housing the metal powder filters is mounted on the mixing chamber plate. Filter characteristic of different metal powder filters at room temperature and 10 mK are shown, (b) and (c), respectively. Black and dark cyan curves correspond to noise floor and through-put after calibration, respectively. The metal powder filters behave different at room temperature, but show very similar attenuation curves at millikelvin temperature.

Now, let us take a closer look at the influence of temperature on the transmission characteristic. In Fig. 4(a), data of the PCB-copper powder filter at different temperatures are plotted. We observe a clear temperature dependence, as for the other filters (not shown here). With decreasing temperature the fall-off, including the resonances, shifts to higher frequencies. The graphs for 10 mK and 77 K show nearly identical behaviour. In Fig. 4(b) we compare the attenuation curves of the PCB- powder filter design with experimental data from refs [9] and [5]. At room temperature the PCB- copper powder filter (wire length 1.2 m) is comparable with a classical copper powder filter with a wire length of 2 m [9]. At low temperature the PCB- copper powder filter reaches -60 dB at 2.3 GHz, whereas the classical copper powder filter from Lukashenko et al. [5] with 4 m wire length reaches -60 dB at 4.2 GHz. In Fig. 4(c) classical stainless steel powder filters from Lukashenko et al. [5] with wire lengths of 4 m and 1.14 m are compared with the PCB- stainless steel powder filter (wire length 1.2 m). Here, the PCB- powder filter has a fall-off at higher frequencies than its classical counterpart.

Lukashenko *et al.* [5] measured the attenuation behaviour at 4 K, whereas Fukushima *et al.* [6] compares measurements at 300 K and 4 K. In agreement with their data, the copper powder filter shows a clear temperature dependence with the fall-off shifted to higher frequencies at low temperature. While the PCB- powder filter shows the same trend for stainless steel, Fukushima *et al.* did not determine a significant temperature dependence. In their work the temperature dependence of the copper powder filter is explained by the reduction of the powder's resistance as the temperature decreases, whereas in the case of stainless steel it remains unclear. Assuming that the reduction of resistance is the reason, the above explanation would be applicable to our copper powder that has almost two orders of magnitude lower resistance at 4 K, assuming a bulk-like residual-resistance ratio (RRR) of ~50. However, it does not explain the temperature dependence of brass and stainless steel (see Fig. 3(c)), which both have a RRR of ~ 1. The real underlying attenuation mechanism is not yet understood.

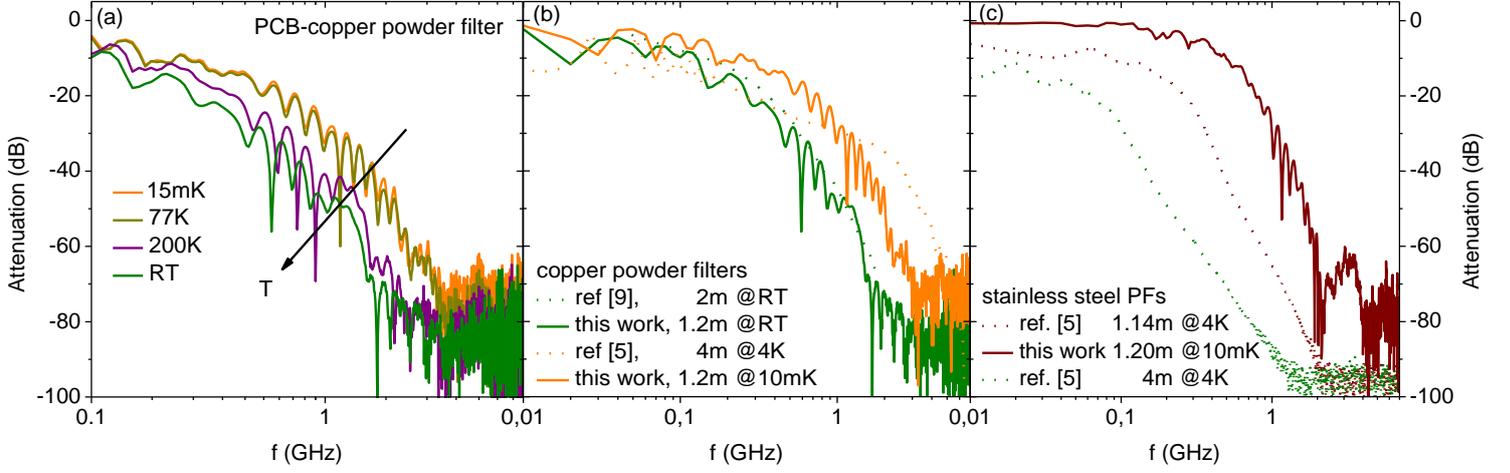

FIG. 4. PCB PFs- temperature dependence and comparison with literature (a): Temperature dependence of PCB-CPF; for decreasing temperature the fall-off is shifted to higher frequencies. Comparison of PCB-PFs (solid line) with classical metal powder filters of different wire length from literature (dashed line), copper (b) and stainless steel (c), respectively.

## IV. EFFECTIVE ELECTRON TEMPERATURE

Finally, we use a single quantum dot in a silicon MOSFET structure [10] to investigate the effect of the copper powder filter on the effective electron temperature $T_e$ of a high-impedance device in a dilution refrigerator. As explained in the introduction the temperature of the electrons in an electrical device can differ from the refrigerator temperature $T_{bath}$. Noise and pick-up of interference are the main origin of elevated electron temperatures. In addition, device specific characteristics have to be taken into account, in our case quantum mechanical level broadening which is determined by the tunnel coupling $\hbar\Gamma$ between the dot and the electron reservoirs. The Coulomb peak of a single-tunneling resonance can then be described by a convolution of a Lorentzian function (tunnel broadening) with the derivative of the Fermi-Dirac distribution function (F-D function; thermal broadening) [11].

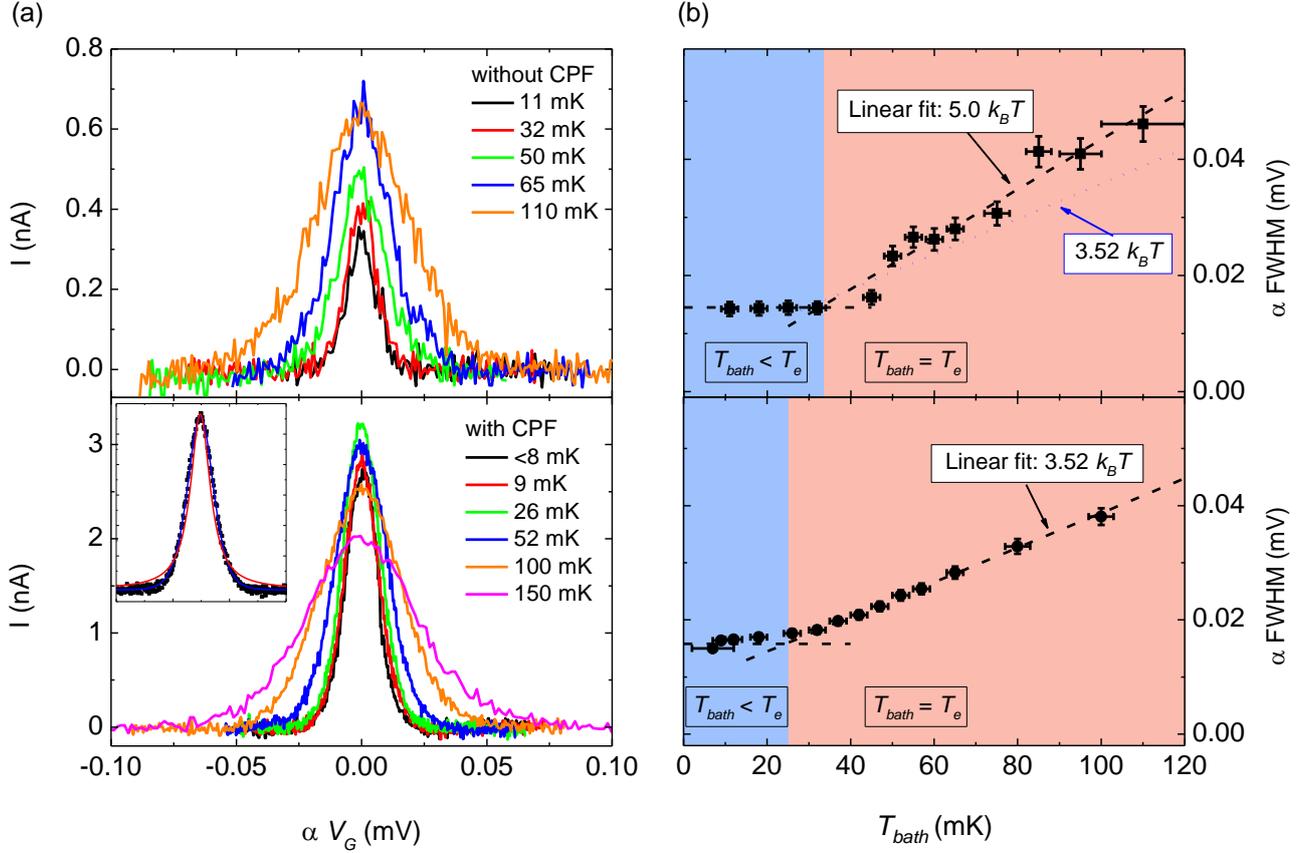

FIG. 5. Effective electron temperature (a): Coulomb peaks at different mixing chamber temperatures without and with CPF, top and bottom, respectively. Inset: Coulomb peak at $T_{bath} = 9$ mK; comparison of peak fitting with a derivative of Fermi-Dirac function (thermal broadening; blue) and Lorentzian function (tunnel broadening; red) (b): peak width versus mixing chamber temperature. In the pink area the the coulomb peak is thermally broadened, meaning that the electron temperature $T_e$ equals the bath temperature $T_{bath}$; in the blue are the electron temperature is higher than the bath temperature.

The measured Coulomb peak increase from base temperature to 500 mK, both with and without mounted copper powder filter, see Fig. 5. By fitting the Coulomb peaks with both Lorentzian (tunnel-broadened) and F-D function (thermally broadened) (inset Fig. 5(a)) we can verify that we operate in a tunnelling regime where the width of the peak is determined by thermal broadening: $\hbar\Gamma \ll k_B T_e$, where $k_B$ is the Boltzmann constant. In Figure 5(b) we plot the FWHM of the peaks versus the mixing chamber temperature $T_{bath}$. Above ~30 mK the FWHM decreases linearly with temperature. Below ~30 mK the FWHM deviates from the proportional trend and remains constant, indicating that in this regime the effective electron temperature is no longer determined by the bath temperature. Similar trends are found with and without copper powder filter.

We examine three methods to extract the electron temperature: method I: Graphical extraction as the inflection point between linear and constant part (graphical); method II: Peak fitting at base temperature with Fermi-Dirac derivative (Fermi-Dirac) and method III: Extrapolation to $T_{bath} = 0$ and evaluation of thermally broadened part (linear fit).

## A. Method I

The first method to extract the electron temperature is to fit the data with a linearly proportional part (thermally broadened, pink areas in Figure 5(b)) and a constant part (broadening at base temperature, pink areas in Figure 5(b)) and extract the effective electron temperature as the interception point. With this method, we get $T_e \approx 35$ mK without CPFs and $T_e \approx 25$ mK with CPFs. This method is quite straightforward and allows a direct graphical extraction of the lowest temperature the electrons can reach. But nevertheless it remains only a rough estimate.

## B. Method II

Method II evaluates the width of the F-D function fitting at base temperature. As mentioned above, the Fermi-Dirac peak shape indicates a thermally broadened quantum mechanical level. Assuming single-level tunnelling [12] the width of the peak relates to the electron temperature as *α FWHM = 3.52$k_B T_e$*, where α is a factor of proportionality relating the applied gate voltage to the potential on the dot. With CPF, α = 0.282 as determined by a F-D function fit at 500 mK, where the electron temperature $T_e$ must equal the bath temperature $T_{bath}$. This value agrees well with the α extracted from the corresponding Coulomb diamond (not shown). The F-D function fit and the linear dependence of the peak width versus temperature with a slope of 3.52$k_B$ confirms the thermally broadened single-level tunnelling regime. Since α is a device specific parameter and is constant at a single Coulomb resonance, the same α is used to extract the effective electron temperature at base temperature. This second method gives an upper bound for $T_e$, since the FWHM is determined by temperature broadening. This upper bound is approximately 47 ± 2 mK with copper powder filters.

## C. Method III

Within the third method we fit the temperature dependence linearly using the α-factor from the peak fit at 500 mK and verify if we have a single-level tunnelling behaviour (slope of 3.52$k_B$). For the measurements with copper powder filter the linear fit gives a slope of 3.52$k_B$ as expected value for single-level tunnelling. Extrapolating this fit to $T_{bath} = 0$ yields a residual peak width of 8.3 μeV. Goldhaber-Gordon *et al.* [1] explain the residual peak width by tunnel broadening, since they describe the peaks as given by a convolution of a Lorentzian of FWHM Γ with the derivative of a Fermi-Dirac function (FWHM of 3.52$k_B T_e$) with a total FWHM of 0.78Γ + 3.52$k_B T_e$ (Γ/h is tunnel rate of electrons from the dot). However, in our case the residual peak width accounts for almost 50% of the FWHM at base temperature, even though the tunnelling rate is only $\hbar\Gamma = 0.022$ μeV as extracted from the fit (inset of Fig. 5(a)). Moreover, the peak can be perfectly fitted with an F-D function and the peak height is much too low for tunnel broadening [11]. Since this offset is not related to thermal broadening we subtract this value from the peak width at base temperature. Using α = 0.29, average from the peak fit at 500 mK and the Coulomb diamond, we get an effective electron temperature of $T_e = 22 \pm 2$ mK. The exact origin of the additional peak broadening cannot be pinpointed, but the peak shape excludes tunnel broadening and the temperature dependence of the FWHM strongly suggests that it does not originate from thermal broadening.

The FWHM shows the same behaviour without CPFs: at higher temperatures, a linear slope is observed, which saturates to a finite peak width at lower temperatures. A linear fit at high temperatures gives a slope of $5k_B$, which deviates from the expected $3.52k_B$. Accounting for this deviating slope, the effective electron temperature is calculated as $T_e = 35 \pm 7$ mK, the higher uncertainty originating from noisier measurements. We note that the measurements were performed in two separate cool-downs and with different quantum dots, which can cause different noise levels and deviating slopes.

The results in Figure 5 show that a combination of low-pass RC filters and Pi-filters results in an effective electron temperature $T_e$ of 35 mK. Inclusion of the PCB- copper powder filters lowers $T_e$ to 22 mK.

TABLE I. Comparison of the methods to extract the effective electron temperature.

| Method | $T_e$ without CPF | $T_e$ with CPF | remarks |
|---|---|---|---|
| I: graphical | ~ 35 mK | ~ 25 mK | Pro: Straight-forward; <br> Con: Fast and rough estimate; |
| II: Fermi-Dirac | < 50 mK | < 47 $\pm$ 2 mK | Pro: Shape of fit indicates broadening type (tunnel/ thermal); <br> Con: only accurate for single-level tunnelling and $\hbar\Gamma \ll k_B T_e$; <br> Gives upper limit for $T_e$ |
| III: linear fit | 35 $\pm$ 7 mK | 22 $\pm$ 2 mK | Pro: With known α, single-level tunnelling (slope of 3.52 $k_B$) can be verified; <br> Con: Does not explain where residual (at zero T) peak broadening comes from, however not necessary for temperature determination |

In conclusion, we have shown a novel and easy way to fabricate metal powder filters with reproducible design using printed circuit boards as basis. We investigated the transmission characteristics of different metal powders. All metal powder filters perform better at room temperature than at cryogenic temperatures. All filters behave nearly the same below 4 K, in contrast to our expectations and results previously found in other systems. According to literature the underlying absorption mechanism is energy loss caused by the skin effect. This mechanism implies a stronger attenuation for a higher metal powder resistivity, while our filters with metal resistivities differing by more than two orders of magnitude attenuate very similar. Finally, we investigated the influence of the copper powder filter on the effective electron temperature. Operating a dilution refrigerator with printed-circuit-board copper powder filters we reach an effective electron temperature of 22 mK in a high-impedance device, closely approximating the base temperature of the cryostat (10 mK).

W.G.v.d.W. acknowledges financial support from the European Research Council, ERC Starting Grant no. 240433.

ASD acknowledges support from the Australian Research Council Centre of Excellence for Quantum Computation and Communication Technology (project number CE110001027) and the US Army Research Office under contract number W911NF-08-1-0527.

FAZ acknowledges support from the Foundation for Fundamental Research on Matter (FOM), which is part of the Netherlands Organization for Scientific Research (NWO), and support from the European Commission under the Marie Curie Intra-European Fellowship Programme.